\documentstyle[sprocl,psfig]{article}

\def\Journal#1#2#3#4{{#1} {\bf #2}, #3 (#4)}

\def\PRL{\em Phys. Rev. Lett.}
\def\PRD{{\em Phys. Rev.} D}

\def\be{\begin{equation}}
\def\ee{\end{equation}}
\def\bea{\begin{eqnarray}}
\def\eea{\end{eqnarray}}
\begin{document}

\title{
A TRANSPORT MODEL FOR HEAVY ION COLLISIONS AT RHIC
}

\author{
BIN ZHANG [1], CHE MING KO [1], BAO-AN LI [2], ZIWEI LIN [1]
}

\address{
[1] Cyclotron Institute and Physics Department, \\
Texas A\&M University, \\
College Station, TX77843-3366, USA
}

\address{
[2] Department of Chemistry and Physics, \\
P.O. Box 419, Arkansas State University, \\
State University, AR 72467-0419, USA
}


\maketitle\abstracts{
To study heavy ion collisions at energies to be available from the
Relativistic Heavy Ion Collider (RHIC), we have developed a
transport model that includes both initial partonic and final
hadronic interactions. Specifically, the parton cascade model ZPC,
which uses as input the parton distribution from the HIJING model,
is extended to include the quark-gluon to hadronic matter
transition and also final-state hadronic interactions based on the
ART model. Comparisons with SPS data and predictions for RHIC are
reported. }

\section{Introduction}
The scheduled beginning of experiments at the Relativistic Heavy
Ion Collider this year will start an exciting new era of nuclear
and particle physics. The estimated high energy density in central
heavy ion collisions at RHIC will lead to the formation of a large
region of deconfined matter of quarks and gluons, the Quark Gluon
Plasma (QGP). This gives us an opportunity to study the properties
of QGP and its transition to the hadronic matter, which would then
shed light on the underlying fundamental theory of strong
interactions, the Quantum Chromodynamics.

To study the interactions of many strongly interacting particles
and to relate the experimental results to the underlying theories,
Monte Carlo event generators are needed. In this talk, we report a
recently developed transport model that starts from initial
conditions that are motivated by the perturbative QCD and
incorporates the subsequent partonic and hadronic space-time
evolution. In particular, we will describe the HIJING model
\cite{hijing1} that is used to generate the initial phase space
distribution of partons and the ZPC model \cite{zpc1} that models
the parton cascade. We will also give a brief discussion of the
modifications made to the default HIJING fragmentation scheme and
ART \cite{art1} hadron evolution. This is followed by some results
and a summary.

\section{Elements of the new transport model}
\subsection{HIJING initial conditions}
HIJING is a Monte-Carlo event generator for hadron-hadron,
hadron-nucleus, and nucleus-nucleus collisions. The main assumption
made in simulating nucleus-nucleus collisions is the binary
approximation in which a nucleus-nucleus collision is decomposed
into binary nucleon-nucleon collisions. For each pair of nucleons,
the impact parameter is calculated using nucleon transverse
positions generated according to a Wood-Saxon nuclear geometry. The
eikonal formalism is then used to determine the probability for a
collision to occur. For a given collision, one further determines
if it is an elastic or an inelastic collision, a soft or hard
inelastic interaction, and the number of jets produced in the hard
interaction. To take into account the nuclear effect in hard
scatterings, an impact parameter dependent parton distribution
function based on the Mueller-Qiu parameterization of the nuclear
shadowing is used. Afterwards, the PYTHIA routines are called to
describe the hard interactions, while the soft interactions are
treated according to the Lund soft momentum transfer model.

After the momentum information of each produced parton is known,
HIJING uses a simple model, in which gluons lose energy to the
wounded nucleons close to their straight line trajectories, to
quench the produced jets (minijets). In the present model, we
replace the quenching in HIJING by the cascade of produced partons.

\subsection{ZPC parton evolution}

The parton cascade in our calculation is carried out with the ZPC
model. At present, this model includes only gluon-gluon elastic
scatterings. To generate the initial phase space distribution, the
formation time for each parton is determined according to a
Lorentzian distribution with a half width $t_f=E/m_T^2$. Positions
of formed partons are calculated by straight line propagation of
the parton from its parent nucleon position. During the time of
formation, partons are considered to be part of the coherent cloud
of their parents and thus do not suffer rescatterings.

The gluon-gluon scattering cross section is taken to be the leading
divergent cross section regulated by a medium generated screening
mass, which is related to the phase space density of produced
partons. In the present study, the simple expression, $\mu^2\approx
3\pi\alpha_S/R_A^2\times(N_G/2Y)$, is used to estimate the
screening mass.

\subsection{Hadronization}
Once partons stop interacting, they are converted into hadrons
using the HIJING fragmentation scheme after an additional proper
time of approximate $1.2$ fm. In the default HIJING fragmentation
scheme, a diquark is treated as a single entity, and this leads to
an average rapidity shift of about one unit. We modify this
fragmentation scheme to allow the formation of diquark-antidiquark
pairs. In addition, the $BM\bar{B}$ probability is $80\%$ among the
produced diquark-antidiquark pairs, while the rest are
$B\bar{B}$'s.

\subsection{ART hadron transport model}

For hadron evolution, we use the ART model, which is a very
successful transport model for heavy ion collisions at AGS
energies. To extend the model for heavy ion collisions at RHIC, we
have included anti-nucleon annihilation channels, the inelastic
interactions of kaons and anti-kaons, and neutral kaon production.
In the ART model, multiparticle production is modeled through the
formation of resonances. Since the inverse double resonance
channels have smaller cross sections than those calculated directly
from the detailed balance, we thus adjust the double resonance
absorption cross sections to improve the model.

\section{Results}

\subsection{comparison with SPS data}

\begin{figure}[ht]
%\rule{5cm}{0.2mm}\hfill\rule{5cm}{0.2mm}
%\vskip 2.5cm
%\rule{5cm}{0.2mm}\hfill\rule{5cm}{0.2mm}
\psfig{file=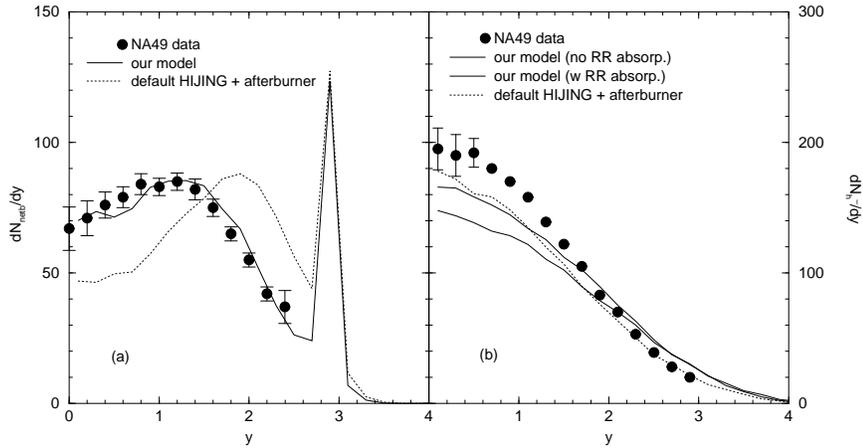,height=2.3in,angle=-90}
\caption{
Comparisons between our model calculation and the NA49 data for (a)
net baryon rapidity distribution and (b) negative hadron rapidity
distribution.
\label{fig:sps}}
\end{figure}

\begin{figure}[ht]
%\rule{5cm}{0.2mm}\hfill\rule{5cm}{0.2mm}
%\vskip 2.5cm
%\rule{5cm}{0.2mm}\hfill\rule{5cm}{0.2mm}
\psfig{file=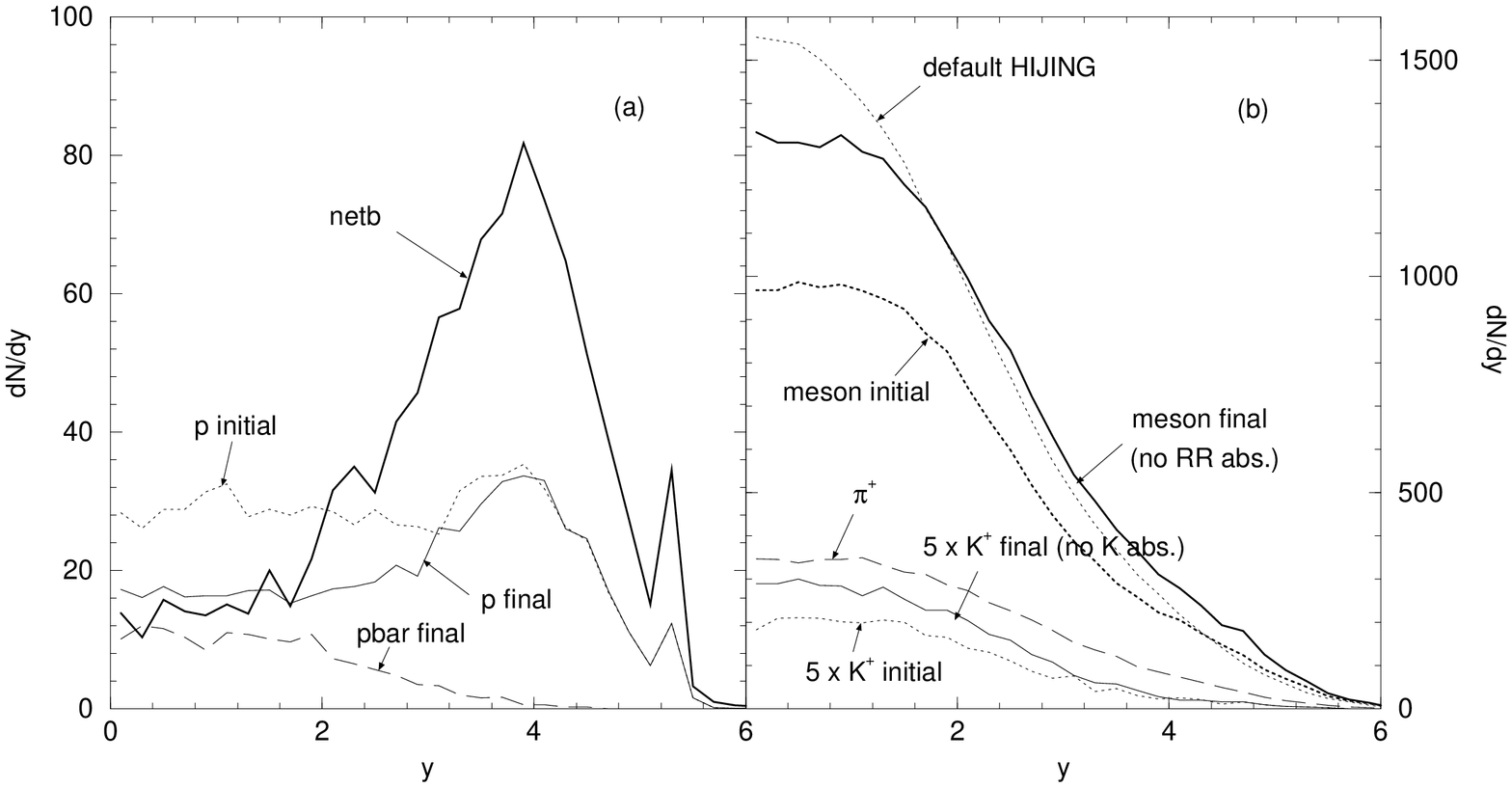,height=2.3in,angle=-90}
\caption{
Predictions of (a) baryon and (b) meson rapidity distributions for
RHIC Au+Au central (b=0) collisions.
\label{fig:rhic}}
\end{figure}

Fig.\ref{fig:sps}(a) shows a comparison of the net baryon rapidity
distribution between our model results and the NA49 data
\cite{na49a} for $5\%$ central Pb(158GeV)+Pb collision. The dotted
curve gives the result using the default HIJING fragmentation
scheme with afterburner. It shows a visible smaller rapidity shift
than that obtained using the modified HIJING fragmentation scheme
with afterburner, shown by the solid curve. The latter is seen to
give a satisfactory description of the data. Fig.\ref{fig:sps}(b)
shows the negative hadron distribution. We see that the
modification to HIJING fragmentation scheme increases the rapidity
width due to the exchange of meson and baryon positions and that
final-state interactions included in the default ART model lowers
the central rapidity density. When one turns off the double
resonance cross sections, the central rapidity density increases to
a reasonable value comparing to the NA49 data \cite{na49a}. An
enhancement of strange particles is also observed but the magnitude
is very sensitive to their formation time.

\subsection{predictions for RHIC}

We show our predictions for RHIC Au+Au central (b=0) collisions in
Fig.\ref{fig:rhic}. The net baryon peak position is similar to the
prediction \cite{brahms1} of Fritiof1.7, but the peak height at 80
is smaller than the 100 predicted by Fritiof1.7. On the other hand,
the central rapidity height is around 15 which is similar to the
Venus4.02 prediction but is much higher than the Fritiof1.7
prediction. We see that many anti-protons still survive after
absorption in the hadronic matter, leading to a value of about 10
at central rapidities. The final meson central rapidity height is
much large than the one at the initial time when many rho mesons
exist. Although the central rapidity height is similar to Venus,
our calculation gives a much wider rapidity distribution. Results
using the default HIJING show a similar distribution except that
the central rapidity density is higher. Including the
afterburner but without double resonances induced pion absorption
also gives a similar meson distribution. Also shown in the figure
is the distribution of kaons produced from both string
fragmentation and hadronic production. The latter is seen to
enhance kaon production significantly.

\section{Summary}

The present study shows that because of the production of
diquark-antidiquark pairs, there is a relatively large rapidity
shift of net baryons comparing to the default HIJING fragmentation
scheme. Many anti-protons survive the final-state interactions and
are expected to be observed at RHIC. Also, our model gives a wider
rapidity plateau at central rapidity than the predictions from
default HIJING model. Furthermore, kaon production is appreciably
enhanced due to production from hadronic interactions. However,
many new elements need to be incorporated into the code, e.g.,
parton inelastic scatterings, dynamical screening, improved
hadronization, etc. Predictions based on such an improved model
will be reported in the future.

%%%%%%%%%%%%%%%%%%%%%%%%%%%%%%%%%%%%%%%%%%%%%%%%%%%%%%%%%%%%

\section*{Acknowledgments}
This work is supported by the NSF Grant PHY-9870038, the Welch
Foundation Grant A-1358, and the Texas Advanced Research Project
FY97 010366-068.

\end{document}